\documentclass{mpe_report}

\usepackage{psfig,graphicx,epsfig}
\usepackage{color}
\usepackage{amsmath,amssymb,epic,eepic,array}

\begin{document}

\pagenumbering{arabic}
\setcounter{page}{169}

 \renewcommand{\FirstPageOfPaper }{169}\renewcommand{\LastPageOfPaper }{172}
\title{Improved methods for modeling pulse shapes of accreting millisecond pulsars}
\author{D.Leahy\inst{1}, S. Morsink\inst{2} \and C.Cadeau\inst{2}}  
\institute{Department of Physics and Astronomy, University of Calgary, Calgary AB, T2N~1N4, Canada
\and  Theoretical Physics Institute, Department of Physics, University of Alberta, Edmonton, AB, T6G~2J1, Canada}

\authorrunning{D.Leahy et al.}
\maketitle

\begin{abstract}
 Raytracing computations for light emitted from the surface of a rapidly rotating neutron star are carried out in order to construct light curves for accreting millisecond pulsars. These calculations are for realistic models of rapidly rotating neutron stars which take into account both the correct exterior metric and the oblate shape of the star. We find that the most important effect, comparing the full raytracing computations with simpler approximations currently in use, arises from the oblate shape of the rotating star. Approximating a rotating neutron star as a sphere introduces serious errors in fitted values of the star's radius and mass if the rotation rate is very large. However, for lower rotation rates acceptable mass and radius values can be obtained using the spherical approximation. 
\end{abstract}

\section{Introduction}

One of the primary problems in neutron star astrophysics is the determination
of the mass-radius relation of neutron stars through observations. This would allow the 
determination of the equation of state (EOS) of cold supernuclear density material. Neutron stars are too small
to allow direct measurements of their radii. 
However, an indirect method for inferring
the radius is through modeling observed light curves for 
X-ray pulsars and neutron stars that exhibit pulsations during
type I X-ray bursts. This method requires
raytracing of photons emitted from the star's surface in order to predict a lightcurve seen by an observer. 
In this paper we present the first raytracing calculations for 
rapidly-rotating neutron stars that include the correct metric and the correct shape of the neutron star. 

In the case of the slowly rotating X-ray pulsars complicated accretion columns are needed 
in order to fit the observations (Leahy 2004), but it is quite possible that the simpler models discussed here, with 
emission from the surface, are adequate for the more 
rapidly-rotating ms X-ray pulsars.
Another important input to light curve models is the spectrum and emissivity of the emitting 
region. For example, for the 2.5 ms X-ray pulsar
SAX~J1808.4-3658, Poutanen \& Gierli\'{n}ski(2003) have shown that a hybrid spectrum including isotropic blackbody and anisotropic 
Comptonized emission is needed.
The assumed shape of the emitting region also strongly affects the pulse profile. 
Most studies have focused on either infinitesimal spots or simple Gaussian profiles. 

The gravitational field outside of the star is important: 
it redshifts the photons and determines their propagation. 
For example, Fig. 1 shows ray paths from the surface of a 
1.4$M_{\odot}$ object to an observer located at $z=-\infty$. Three
cases for the stellar radius are shown, with the smallest corresponding
to the limiting case of the observer being just able to see the 
whole neutron star surface.
Fig. 1 illustrates that the visibility, and hence light curve, 
that an observer sees from an emission region on the surface is
strongly affected by light bending.
For non-rotating stars the 
Schwarzschild metric is sufficient to determine ray paths and redshift
 (Pechenick, Ftaclas, \& Cohen 1983).
The Schwarzschild + Doppler (S+D) approximation (Miller \& Lamb 1998)
has been introduced for calculating light curves of slowly 
rotating stars: 
the gravitational field is modeled by the Schwarzschild metric and the
rotational effects are approximated by using the special relativistic Doppler transformations.
For rapidly rotating neutron stars, the metric must be computed numerically. 
The Kerr black hole metric can be used as an approximation to a
rotating neutron star metric, which has led to the 
Spherical Kerr approximation (SK) where a spherical
surface is embedded in the Kerr spacetime (Braje, Romani, \& Rauch 2000). 

Another important input is the shape of the star's surface,
which for a rotating fluid star is an oblate spheroid.
In almost all previous calculations it has been assumed that the surface of the star is a sphere.  The 
oblate shape of the star has two main effects: i) the gravitational 
field near the the poles is stronger than near the equator, so the redshift and photon deflection
are larger for photons emitted near the poles; ii)
light is emitted at different 
angles with respect to the normal to the surface if the surface is an oblate spheroid instead of a sphere.  As a result some parts of a spherical star that an observer can't see will be
visible if the star is actually oblate. Also some parts of a spherical
star that an observer can see will not be visible for the
oblate star. 
The second effect above is the most important, and leads to failure of the S+D and SK
approximations for rapid rotation, as is illustrated below.  

In this paper we present light curve calculations for rapidly
rotating neutron stars using the exact
neutron star metric and include the correct shape of the oblate
neutron star. We also compare these calculations to various
approximations, in particular the Schwarzchild plus doppler and
spherical Kerr approximation. The errors that can occur
in derived mass and radius by using the Schwarzchild plus doppler
approximation are discussed.
In a previous paper (Cadeau, Leahy, \& Morsink 2005) (CLM) 
similar calculations for rapidly rotating neutron stars
were carried out that were restricted to the rotational equatorial plane of the neutron star.

\section{Calculation of light curves}

\begin{figure}
\centerline{\psfig{file=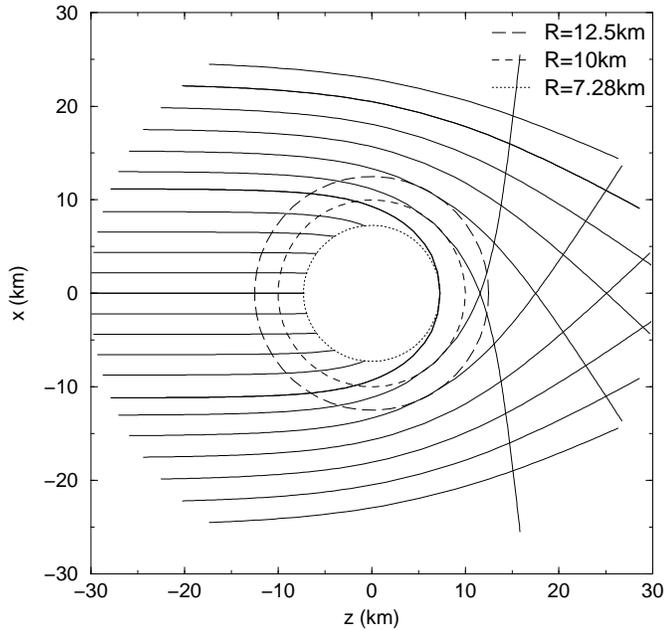,width=8.8cm,clip=} }
\caption{Ray paths from the surface of a 1.4$M_{odot}$ compact spherical object that
reach an observer at $z=-\infty$, for three different radii of the
object.
}
\end{figure}

\begin{figure}
\centerline{\psfig{file=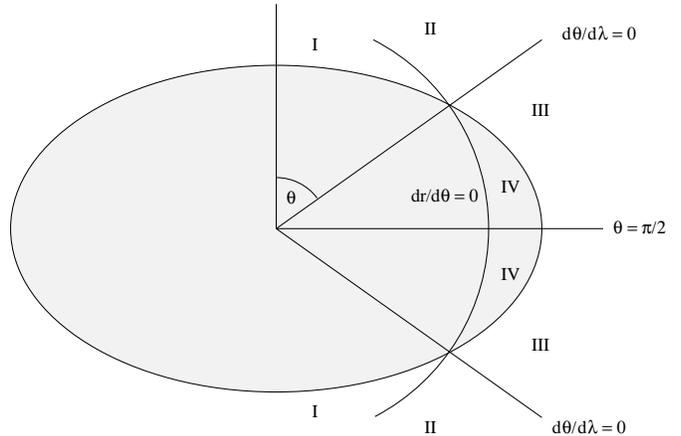,width=8.8cm,clip=} }
\caption{Photons emitted from a point on the star at an angle $\theta$ from
the spin axis can be emitted into regions 
I, II, or III if the star's surface is an oblate spheroid. If the star's surface is
spherical, photons can only be emitted into regions II, III, or IV.
}
\end{figure}

The inputs to the calculations are briefly described here.
We assume the photons are emitted from the neutron star surface,
with isotropic emissivity and from an infinitessimal spot.
We can calculate an arbitrarily shaped spot and arbitrary emissivity
but use the above case here for simplicity.
The algorithm of Cook, Shapiro, \& Teukolsky(1994) is used to
calculate the rotating neutron star metric. 
Two equations of state (EOS) are used from the Arnett \& Bowers(1977) catalog which 
span a range of stiffness:  EOS A is one of the softest EOS and 
EOS L is one of the stiffest. We present results for neutron 
stars with masses of $1.4 M_\odot$ and spin frequencies from 
100 to 600 Hz in 100 Hz increments
in order to explore the spin dependent effects. 

Details of the calculations, including the raytracing along
geodesics, are given in Cadeau et al. (2006). 
We emphasize the importance of 
stellar oblateness on the raytracing here.
Fig. 2 is a schematic view of the cross-section of the star including
the polar axis, showing both the spherical case and oblate case
(the oblateness is exaggerated for clarity).
Consider an emitting point on the star  near $\theta=45^\circ$
as indicated. If the surface is oblate, photons can only be
emitted into regions I, II, or III. In contrast if the surface
is spherical, photons can only be emitted into regions II, III, 
or IV. Thus the spherical approximation erroneously includes photons
propagating to the observer in region IV and omits photons 
propagating to the observer in region I. 

\section{Results}
\subsection{Comparison of exact and approximate methods}
\begin{figure}
\centerline{\psfig{file=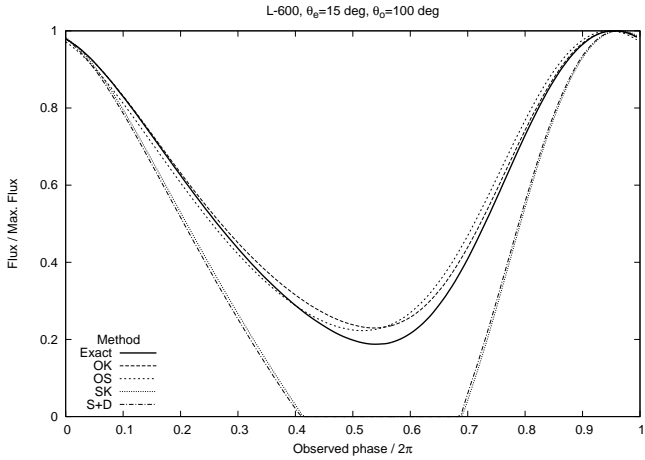,width=8.8cm,clip=} }
\caption{Light curves for a 600 Hz, 1.4 $M_\odot$ neutron star with EOS L, emission from
an angle of $15^\circ$ from the North pole and an observer at an inclination angle of $100^\circ$
from the North pole.
}
\end{figure}

\begin{figure}
\centerline{\psfig{file=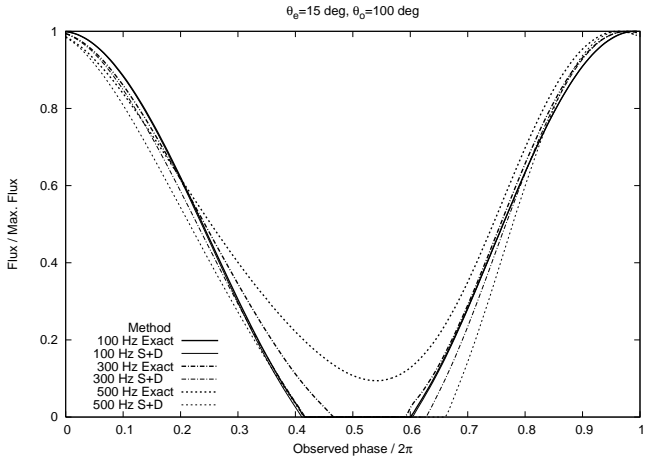,width=8.8cm,clip=} }
\caption{Light curves for 1.4 $M_\odot$ neutron stars with EOS L, emission from
an angle of $15^\circ$ from the North pole and an observer at an inclination angle of $100^\circ$
from the North pole.  Light curves for spin rates of 100, 300 and 500 Hz are shown. Exact light 
curves are illustrated with bold curves. For comparison the approximate Schwarzschild + Doppler (S+D)
light curves are also shown. The S+D and exact light curves for the 100 Hz star overlap, so the S+D curve
can't be distinguished from the exact in this figure.
}
\end{figure}

We compare light curves calculated using the exact numerical 
methods with those calculated using the Schwarzschild plus Doppler
(S+D) and spherical Kerr (SK) approximations.
We also introduce the oblate Schwarzschild (OS) and oblate Kerr (OK) 
approximations, in which the shape of the star is calculated
using the exact neutron star metric and the raytracing is done
using either Schwarzschild metric or Kerr metric, respectively 
(instead of the exact neutron star metric).
Fig. 3 shows light curves 
for the stiff, large radius star (EOS L) rotating
at 600 Hz, with a hot spot at 15$^\circ$ for the north rotation
pole and the observer located at 100$^\circ$ from the north pole.
Both S+D and SK calculations give light curves with a significant
eclipse of the hotspot, whereas the OS, OK and exact calculations 
give light curves without any eclipse. This difference is mainly
due to the different visibility by the observer of the neutron
star surface between spherical and oblate cases, as illustrated
in Figure 1.
We also note that both OS
and OK light curves are slightly different that the light curve
from the exact calculation.

Figure 4 illustrates the effect of rotation rate on the light curves
for the S+D approximate calculations and for the exact calculation.
The hot spot is at an angle of $15^\circ$ from the north
rotation pole and the observer is at $100^\circ$. The different
neutron stars are all with the stiff EOS L, and rotating at frequencies 100, 300 and 500 Hz. 
For 100 Hz rotation rate
the exact (bold) and approximate S+D (light) curves are almost identical. Thus for 100 Hz, where the rotation rate is slow and 
the oblateness is small, the S+D light curve is an excellent approximation to the exact light curve. 
The light curves
for 300 Hz rotation rate (dot-dashed lines) show a significant difference between the exact (bold) and
S+D (light) curves. In both cases the hot spot is eclipsed, 
but for the exact light curve the 
eclipse lasts half as long as when calculated using the
 S+D approximation.  In the case of 500 Hz (dotted lines) the exact light curve
(bold) has no eclipse while the S+D approximation (light) does, similar to the case of 600 Hz rotation rate shown in Fig. 3.

\subsection{Fitting exact light curves using approximate methods}

Existing methods for fitting the  mass and radius from light curves
all approximate the neutron star surface as a sphere. The goal is
to determine the difference between mass and
radius derived from fitting a light curve generated with 
the spherical approximation to an exact light curve. 
First, exact light curves 
are created using the above described numerical calculation, 
including ray tracing in the exact metric and correct shape of the neutron star surface. This was done for several
different neutron stars models, emission latitudes and observer inclination angles. Then 
a S+D fitting program was used to deduce the mass and radius of the neutron star and the emission-observer
geometry from the pulse shape. Only the neutron star spin frequency
was known prior to the fitting. 
For simplicity, both the exact calculation and the fitting program
assumed infinitesimal hot spot size and isotropic emissivity.

Exact neutron star metrics were generated for rotation frequencies 
of 100, 200, 300, 400, 500 and 600 Hz and for the two equations of
state mentioned above (EOS A and EOS L). For each of the 12 cases
exact light curves were generated for several sets of hot spot
angle with respect to the rotation axis (hot spot latitude) 
and of observer viewing angle from the rotation axis. 
An S+D fitting program was developed and tested,
then used to fit the light curves from the exact calculation. 
Input parameters (neutron star mass and radius, hot spot and observer
angle) and output parameters were compared. 

Several systematic trends in the fitted vs. input parameters were
found (a detailed comparison is given in Cadeau et al. 2006). We note
that the S+D program is fitting the neutron star radius at the hot
spot location, so we compared the actual neutron star radius at the
latitude of the hot spot with the fitted radius. 
The first trend we found was expected: generally if the hot spot
angle and the observer angle were too small (less than $\sim45^\circ$)
then the exact light curve was very close to sinusoidal. This
resulted in a degenerate set of parameters which could fit the 
light curve equally well. 
In this (degenerate) case one cannot derive unique mass and radius
values from the light curve fitting. Thus below we discuss fits
to cases which had larger hot spot and observer angles and which
produced unique fits to the light curves.

For all rotation rates (where the fits were not degenerate), the S+D
program produced the correct compactness, to better than $\simeq10\%$, 
of the actual neutron star as measured by M/R. 
For rotation rates below about 300 Hz, the 
difference between fitted radius and true radius was less than 
$simeq10\%$, and above 300 Hz the difference increased until, for some
cases at 600 Hz rotation rate, the difference reached $\simeq50\%$.
The difference between fitted mass and true mass was small at low
rotation rate, usually less than $\simeq10\%$ for rotation rates below
300 Hz, but increased with rotation rate.
Generally for EOS A (the soft EOS which resulted in a quite compact neutron star)
the errors in mass and radius were less than for EOS L. This is due
to the significantly larger oblateness for neutron stars with
EOS L than for stars with EOS A.
For the hot spot angle and observer angle the errors were larger:
for all rotation rates the angles could be up to $\simeq50\%$ 
different than the input angles. However in cases where the
light curve had strong non-sinusoidal content (either showing
eclipse of the hot spot or near eclipse) the errors in hot spot
and observer angles were small ($<10\%$).

\section{Summary}

Raytracing computations for light emitted from the surface of a rapidly rotating neutron star were carried out in order to construct light curves for accreting millisecond pulsars. These calculations 
use the full neutron star metric which takes into account both the correct exterior metric, necessary for correct raytracing, and the correct oblate shape of the star. The calculations show that the 
most important effect, in comparing the exact computations with 
simpler approximations such as Schwarzschild plus doppler (S+D) or
spherical Kerr (SK), arises from the oblate shape of the rotating star. Approximating a rapidly rotating neutron star as a sphere introduces serious errors in values of the star's radius and mass which are derived from fitting observed light curves. 
However, for rotation rates below about 200 Hz, acceptable mass and radius values can be obtained using the spherical approximation. 
The calculations also show: when the rotation rate is rapid enough that
the S+D approximation fails, the SK approximation also fails.
Thus using the SK approximation is not necessarily better than using
the S+D approximation at any rotation rate. 

The full raytracing calculations are too time-consuming to be used for fitting observational data. Pulse shape fitting algorithms have been developed which include special relativistic effects, light-bending effects due to general relativity, and time-delays. Currently under development is inclusion of oblateness of the neutron star. 
            
\begin{acknowledgements}
We gratefully acknowledge support by Natural Sciences and Engineering
Research Council of Canada.
\end{acknowledgements}
   


              \clearpage

\end{document}